\documentstyle [12pt] {article}
\makeatletter

\@addtoreset{equation}{section}

\ifcase \@ptsize
\or 
\or 
\fi
\advance\textheight by \topskip

\ifcase \@ptsize
\or 
\or 
\fi

\def\@citex[#1]#2{%
\if@filesw \immediate \write \@auxout {\string \citation {#2}}\fi
\@tempcntb\m@ne \let\@h@ld\relax \def\@citea{}%
\@cite{%
  \@for \@citeb:=#2\do {%
    \@ifundefined {b@\@citeb}%
      {\@h@ld\@citea\@tempcntb\m@ne{\bf ?}%
      \@warning {Citation `\@citeb ' on page \thepage \space undefined}}%
      {\@tempcnta\@tempcntb \advance\@tempcnta\@ne%
      \@tempcntb\number\csname b@\@citeb \endcsname \relax%
      \ifnum\@tempcnta=\@tempcntb 
        \ifx\@h@ld\relax%
          \edef \@h@ld{\@citea\csname b@\@citeb\endcsname}%
        \else%
          \edef\@h@ld{\ifmmode{-}\else--\fi\csname b@\@citeb\endcsname}%
        \fi%
      \else
        \@h@ld\@citea\csname b@\@citeb \endcsname%
        \let\@h@ld\relax%
      \fi}%
    \def\@citea{,\penalty\@highpenalty\,}%
  }\@h@ld%
}{#1}}

\makeatother
\begin{document}
\hfuzz=100pt
\textheight 24.0cm
\topmargin -0.5in
%
%
%
%
\newcommand{\be}{\begin{equation}}
\newcommand{\ee}{\end{equation}}
\newcommand{\bea}{\begin{eqnarray}}
\newcommand{\eea}{\end{eqnarray}}
\begin{titlepage}
\makeatletter
\def \thefootnote {\fnsymbol {footnote}} \def \@makefnmark {
\hbox to 0pt{$^{\@thefnmark }$\hss }}
\makeatother
\begin{flushright}
BONN-HE-93-51\\
December, 1993\\
hep-th/9312182
\end{flushright}
\vspace{1.5cm}
\begin{center}
{ \large \bf On Bosonic and Supersymmetric Current Algebras\\
             for Non-semi-simple Groups}\\
\vspace{2cm}
{\large\bf Noureddine Mohammedi}
\footnote{e-mail: nouri@avzw02.physik.uni-bonn.de}
\footnote
{Work supported by the Alexander von Humboldt-Stiftung.}
\\
\vspace{.5cm}
Physikalisches Institut\\
der Universit\"at Bonn\\
Nussallee 12\\ D-53115 Bonn, Germany\\

\baselineskip 18pt
\vspace{1cm}
{\large\bf Abstract}
\end{center}
We present a systematic approach to constructing current
algebras based on non-semi-simple groups. The Virasoro central
charges corresponding to these current algebras are not, in general,
given by integer numbers. The key point in this construction is
that the bilinear form appearing in the current algebra can be
different from the bilinear form used to raise and lower group
indices. The action which realises this current algebra as its
symmetry is also found.  \\
\setcounter {footnote}{0}
\end{titlepage}
\baselineskip 20pt
\section{Introduction}

Recently, a Wess-Zumino-Novikov-Witten (WZNW) model describing
string propagation on a four dimensional Lorentz-signature
space-time was built on the centrally extended Euclidean group
in two dimensions [1]. {}Further studies of this model [2,3] and
generalisations to higher dimensions were  presented in [4,5].
\par
All these models describe string backgrounds with a target space
metric having a covariantly constant null Killing vector [6-10].
Another feature of the WZNW models
constructed so far [1,4,5], is that they all have integer central charges.
The central charges
are equal to the dimensions of the group manifold built
on the non-semi-simple groups.
\par
In this note, we systematically
describe both the bosonic and the supersymmetric current algebras
based on non-semi-simple groups.
The Virasoro central charges corresponding to these current algebras
are not necessarily integer numbers. We take the bilinear form
entering the operator product expansions of the current algebra
to be different from the bilinear form which raises and
lowers group indices. We then construct a Virasoro generator
via the Sugawara method. The generators of the current algebra
are primary fields of conformal weight (1,1) with respect to
this energy-momentum tensor. {}Finally, we present an action which
has the current algebra as its symmetry.

\section{The Bosonic Current Algebra}

Let $\cal G$ be a non-semi-simple group whose Lie algebra is given by
\be
\left[T^a\,,\,T^b\right]=f^{ab}_{\,\,\,\,\,\,c}T^c\,\,\,.
\ee
The Cartan bilinear form, $\gamma^{ab}$, defined by
\be
\gamma^{ab}=f^{ea}_{\,\,\,\,\,\,c}f^{cb}_{\,\,\,\,\,\,e}
\ee
is degenerate (not invertible) for non-semi-simple groups
and cannot, therefore, be used to raise and lower group
indices. However, $\gamma^{ab}$ is an invariant of the
group, namely
\be
f^{ab}_{\,\,\,\,\,\,c}\gamma^{cd}+
f^{ad}_{\,\,\,\,\,\,c}\gamma^{cb}=0\,\,\,.
\ee
\par
In order to be able to raise and lower group indices,
we need to look
for another bilinear form, denoted here by $\Omega^{ab}$,
which is symmetric, invertible and is an invariant of the
group. In other words, we must have
\bea
&f^{ab}_{\,\,\,\,\,\,c}\Omega^{cd}+
f^{ad}_{\,\,\,\,\,\,c}\Omega^{cb}=0&\nonumber\\
&\Omega^{ab}\Omega_{bc}=\delta^a_c\,\,\,\,,\,\,\,\,\,
\Omega^{ab}=\Omega^{ba}\,\,\,,
\eea
where $\Omega_{ab}$ is the inverse matrix of $\Omega^{ab}$.
In what follows, we will always assume that such a bilinear form
does exist.
\par
The bosonic current algebra built on this Lie algebra is
generated by the holomorphic  currents $J^a(z)$ having
the operator product expansion
\be
J^a(z)J^b(w)=-k{g^{ab}\over {(z-w)^2}}+f^{ab}_{\,\,\,\,\,\,c}
{J^c(w)\over{(z-w)}}\,\,\,\,.
\ee
The new symmetric matrix $g^{ab}=g^{ba}$ is a new bilinear form that
we are going to determine.
In fact, the associativity of the above operator product
expansions shows that $g^{ab}$ is an invariant of the group obeying
\be
f^{ab}_{\,\,\,\,\,\,c}g^{cd}+
f^{ad}_{\,\,\,\,\,\,c}g^{cb}=0\,\,\,.
\ee
Therefore $g^{ab}$ must be a linear combination of $\gamma^{ab}$
and $\Omega^{ab}$
\be
g^{ab}=m\gamma^{ab}+n\Omega^{ab}\,\,\,,
\ee
where $m$ and $n$ are two constants that we are going
to find. Here $g^{ab}$ need not be invertible.
\par
The energy-momentum tensor is then written as a general
quadratic sum of the currents $J^a(z)$
\be
T(z)=L_{ab}:J^aJ^b:(z)\,\,\,\,,
\ee
with $L_{ab}$ a symmetric matrix. We will determine $L_{ab}$
by requiring that the currents $J^a(z)$ are primary fields
of conformal weight 1 with respect to the stress tensor $T(z)$.
Therefore by requiring that
\be
T(z)J^a(w)={J^a(w)\over{(z-w)^2}}+
{\partial J^a(w)\over{(z-w)}}
\ee
we get the following equations for the matrices
$L_{ab}$ and $g^{ab}$ {\footnote {
$L_{ab}$ would have to satisfy a master equation if we did not
require $J^a(z)$ to be of conformal weight equal to one [11,12].}}
\bea
&L_{cb}f^{ba}_{\,\,\,\,\,\,e}+L_{eb}f^{ba}_{\,\,\,\,\,\,c}=0&
\nonumber\\
&-2kL_{cb}g^{ba}+L_{bd}f^{ab}_{\,\,\,\,\,\,e}f^{ed}_{\,\,\,\,\,\,c}
=\delta^a_c&\,\,\,.
\eea
The first equation of this set is equivalent to the first equation in
(2.4) and is uniquely solved by
\be
L_{ab}=l\Omega_{ab}\,\,\,\,,
\ee
where $l$ is any arbitrary number.
Using this solution, the second equation of the set leads to
\be
g^{ab}=-{1\over {2kl}}\left(\Omega^{ab}-l\gamma^{ab}\right)\,\,\,.
\ee
\par
With these expressions for $L_{ab}$ and $g^{ab}$, the above
energy-momentum tensor satisfies the Virasoro algebra
\be
T(z)T(w)={c\over {2(z-w)^4}}+{2T(w)\over {(z-w)^2}}
+{\partial T(w)\over {(z-w)}}\,\,\,.
\ee
The central charge of this algebra is given by
\be
c=\dim({\cal G}) - l\gamma^{ab}\Omega_{ab}\,\,\,,
\ee
where $\dim({\cal G})$ is the dimension of the non-semi-simple
group ${\cal G}$
\par
In performing the above operator product expansion, we have defined
the normal ordered product of two operators $A(z)$ and $B(z)$
at coincident points as [13]
\be
:AB:(z)={1\over 2\pi i}\oint_{C_{z}}
{{\rm d}x\over x-z}A(x)B(z)\,\,\,\,,
\ee
where the countour of integration, $C_{z}$,
surounds the point $z$. This definition
leads to the following form of
Wick's theorem for calculating the product
expansion of $A(z)$ with a composite field $:BC:(w)$
\be
\underbrace{A(z):BC:(w)}={1\over 2\pi i}
\oint_{C_{w}}{{\rm d}x\over (x-w)}
\{\underbrace{A(z)B(x)}C(w)
+(-1)^{BC}\underbrace{A(z)C(w)}B(x)\}\,\,\,,
\ee
where $(-1)^{BC}=-1$ iff both $B$ and $C$ are fermionic fields and
the contraction $(\,\,\underbrace{\,\,\,\,\,\,\,\,\,}\,\,)$
stands for the singular part in the expansion of the product of
two operators at distinct points.

\section{The Supersymmetric Current Algebra}

Let us now turn our attention to the supersymmetric case
of the above current algebra.
The supercurrent algebra is given by the operator
product expansions of the supercurrents ${\bf J}^a(Z)$
[14,15,16]
\be
{\bf {J}}^a(Z_1){\bf {J}}^b(Z_2)=
-kh^{ab}Z_{12}^{-1}+f^{ab}_{\,\,\,\,\,\,c}Z_{12}^{-1/2}
{\bf J}^c(Z_2)\,\,\,.
\ee
Here $h^{ab}$ is a symmetric matrix and, by associativity
of the super operator product expansions, is an invariant
of the group ${\cal G}$.
In these expressions $Z=(z,\theta)$ denotes the holomorphic
coordinate of two-dimensional superspace and
the symbol $Z_{ij}^M$, $M \in {\cal Z}$,  is defined by
\be
Z_{ij}^M=\left\{\begin{array}{ll}
(z_i-z_j-\theta_i\theta_j)^M\,\,\,, & M\in {\cal Z}\\
(\theta_i-\theta_j)(z_i-z_j-\theta_i\theta_j)^{M-1/2}\,\,\,,
& M\in {\cal Z}+{1\over 2}\end{array}\right.\,\,\,.
\ee
We postulate the super energy-momentum tensor to have the form
\be
{\bf T}(Z)=N_{ab}:D{\bf J}^a{\bf J}^b:(Z)
+M_{abc}:{\bf J}^a:{\bf J}^b{\bf J}^c::(Z)\,\,\,,
\ee
where $N_{ab}$ is symmetric and $M_{abc}$ is totally antisymmetric.
The super covariant derivatives is $D={\partial\over
{\partial\theta}}+\theta{\partial\over{\partial z}}$ obeying
$D^2=\partial$.
\par
These two tensors are then determined by requiring that the
supercurrents ${\bf J}^a(Z)$ are primary operators of dimension
${1\over 2}$ with respect to the super stress tensor and that
${\bf T}(Z)$ satisfies the super Virasoro algebra. Demanding
that
\be
{\bf T}(Z_1){\bf J}^a(Z_2)=
{1\over 2}Z_{12}^{-3/2}{\bf J}^a(Z_2)
+{1\over 2}Z_{12}^{-1}D{\bf J}^a(Z_2)
+Z_{12}^{-1/2}\partial{\bf J}^a(Z_2)
\ee
leads to the following six equations {\footnote{
If we did not demand that ${\bf J}^a(Z)$ is a
primary superfield of conformal weight ${1\over 2}$,
then $N_{ab}$ and $M_{abc}$ would have to satisfy
a set of super master equations [17].}}
\bea
&-kN_{bc}h^{ac}={1\over 2}\delta^a_c&
\nonumber\\
&N_{bc}f^{ab}_{\,\,\,\,\,\,d}-3M_{bdc}h^{ab}=0&
\nonumber\\
&N_{bc}f^{ab}_{\,\,\,\,\,\,d}+N_{bd}f^{ab}_{\,\,\,\,\,\,c}=0&
\nonumber\\
&-kh^{dc}f^{ab}_{\,\,\,\,\,\,d}N_{cb}=0&
\nonumber\\
&M_{bcd}f^{ab}_{\,\,\,\,\,\,e}
+M_{ebd}f^{ab}_{\,\,\,\,\,\,c}
+M_{ecb}f^{ab}_{\,\,\,\,\,\,d}=0&
\nonumber\\
&-kN_{bc}h^{ab}+N_{be}f^{ab}_{\,\,\,\,\,\,d}
f^{de}_{\,\,\,\,\,\,c}-3kM_{bdc}f^{ab}_{\,\,\,\,\,\,e}
h^{ed}={1\over 2}\delta^a_c&\,\,\,\,.
\eea
The first three equations of this set
are uniquely solved by
\be
N_{ab}=l\Omega_{ab}\,\,\,,\,\,\,
h^{ab}=-{1\over 2kl}\Omega^{ab}\,\,\,,\,\,\,
M_{abc}={2l^2\over 3}f_{abc}
\equiv {2l^2\over 3}\Omega_{ar}\Omega_{bs}
 f^{rs}_{\,\,\,\,\,\,c}\,\,\,.
\ee
Here $l$ is again an arbitrary number. The last three
equations in the set are then automatically satisfied
once the Jacobi identities for the structure constants
$f^{ab}_{\,\,\,\,\,\,c}$ are used.
\par
Using these expressions for $N_{ab}$, $M_{abc}$
and $h^{ab}$, the
super energy-momentum tensor does indeed satisfy the
super Virasoro algebra
\be
{\bf T}(Z_1){\bf T}(Z_2)={c\over 6}Z_{12}^{-3}
+\left[{3\over 2}Z_{12}^{-3/2}+
{1\over 2}Z_{12}^{-1}D +Z_{12}^{-1/2}\partial\right]{\bf T}(Z_2)
\,\,\,.
\ee
The central charge of this algebra is given by
\be
c={3\over 2}\dim({\cal G})-l\gamma^{ab}\Omega_{ab}\,\,\,.
\ee
In calculating the operator product expansions, we have used the the
following definition of normal ordering of superfields [13,15]
\be
:A(Z_1)B(Z_1):={1\over 2\pi i}\oint_{C_1}dz\oint d\theta
Z_{12}^{-1/2} A(Z_2)B(Z_1)\,\,\,,
\ee
with the $z_2$-integration contour $C_1$ encircling the point $z_1$.
This normal ordering obeys also the analogue of Wick's theorem
for calculating the operator product expansion of $A(Z)$ with
a composite operator :BC:(W) [13,15]
\bea
\underbrace{:A(Z_1):BC:(Z_2):}
&=&{1\over 2\pi}\oint_{C_2}dz\oint d\theta
Z_{23}^{-1/2}\nonumber\\
&\times &\left\{\underbrace{
A(Z_1)B(Z_3)}C(Z_2)+(-1)^{BC}\underbrace{
A(Z_1)C(Z_2)}B(Z_3)
\right\}\,,
\eea
where $(-1)^{BC}=-1$ iff both $B$ and $C$ are fermion fields.
\par
Let us now examine the supercurrent algebra in components.
The superfields ${\bf J}^a(Z)$ and ${\bf T}(Z)$ can be decomposed
into components as
\bea
{\bf J}^a(Z)&=&\psi ^a(z)+\theta J^a(z)\nonumber\\
{\bf T}(Z)&=&{1\over 2}G(z)+\theta T(z)\,\,\,\,.
\eea
In terms of these components, the supercurrent algebra in (3.1)
yields
\bea
\psi^a(z_1)\psi^b(z_2)&=&-k{h^{ab}\over {(z_1-z_2)}}\nonumber\\
J^a(z_1)\psi^b(z_2)&=&f^{ab}_{\,\,\,\,\,\,c}{\psi^c(z_2)\over
{(z_1-z_2)}}\nonumber\\
J^a(z_1)J^b(z_2)&=&-{k}{h^{ab}\over
{(z_1-z_2)^2}}+f^{ab}_{\,\,\,\,\,\,c}{J^c(z_2)
\over{(z_1-z_2)}}\,\,\,\,.
\eea
\par
The supercurrent algebra can be made to look like a direct
sum of a bosonic current algebra and an algebra of free
Majorana fermions. This can be achieved by introducing a modified
current $\widehat{J}^a(z)$ defined by
\be
\widehat{J}^a(z)=J^a(z)+R^a_{\,\,\,\,bc}:\psi^b\psi^c:(z)\,\,\,\,,
\ee
where $R^a_{\,\,\,\,bc}$ is antisymmetric in the indices
$b$ and $c$, and is determined by requiring that the fermions
$\psi^a(z)$ and the bosonic currents $\widehat {J}^a$
decouple. Indeed, the following operator product expansions
\bea
\psi^a(z_1)\psi^b(z_2)&=&-k{h^{ab}\over {(z_1-z_2)}}
\nonumber\\
\widehat{J}^a(z_1)\psi^b(z_2)&=&0
\nonumber\\
\widehat{J}^a(z_1)\widehat{J}^b(z_2)&=&-k{\left(h^{ab}+{1\over 2k}
\gamma^{ab}\right)\over (z_1-z_2)^2}
+f^{ab}_{\,\,\,\,\,\,c}{\widehat{J}^c(z_2)
\over{(z_1-z_2)}}
\eea
hold only if the tensor $R^a_{\,\,\,\,bc}$ satisfies
\be
h^{bc}R^a_{\,\,\,\,be}=-{1\over 2k}f^{ac}_{\,\,\,\,\,\,e}
\,\,\,\,,\,\,\,\,
h^{ab}=-{1\over 2kl}\Omega^{ab}\,\,\,.
\ee
Notice that we have the relation
\be
h^{ab}+{1\over 2k}\gamma^{ab}=g^{ab}\,\,\,,
\ee
where $g^{ab}$ is the bilinear form previously
found for the bosonic current algebra.
The components of the super stress tensor are written in terms
of these modified currents in the form
\bea
T(z)&=&l\Omega_{ab}
:\widehat{J}^a\widehat{J}^b:(z)-
l\Omega_{ab}:\psi^a\partial\psi^b:(z)\nonumber\\
G(z)&=&2l\Omega_{ab}:\psi^a\widehat{J}^b:(z)-{2l^2\over 3}f_{abc}
:\psi^a:\psi^b\psi^c::(z)\,\,\,\,.
\eea
The components $T(z)$ and $G(z)$ satisfy the usual $N=1$
superconformal algebra. The contribution to the central charge
due to the stress tensor of
the bosonic currents $\widehat{J}^a(z)$ is
$\left[\dim({\cal G})-\gamma^{ab}\Omega_{ab}\right]$
while that of the
fermions $\psi^a(z)$ is given by ${1\over 2}\dim({\cal G})$.
These two contributions add up to give, as expected, the central
charge for the $N=1$ algebra calculated in (3.8).

\section{The WZNW Action}

Let us now find a bosonic action (the supersymmetric action is then
straightforward) having the bosonic current algebra defined in
(2.5) as a symmetry. This action is constructed out of the
``gauge fields" $A_{\mu a}$, defined for an element $g$ in the
Lie group via
\be
A_{\mu a}T^a=g^{-1}\partial_\mu g\,\,\,.
\ee
Under an infinitesimal transformation of the form
\be
g\rightarrow g + \bar\omega g + g\omega\,\,\,,
\ee
where $\omega=\omega_a T^a$ and $\bar\omega=\bar\omega_a T^a$,
the variation of the gauge field $A_{\mu a}$ is given by
\be
A_{\mu a}\rightarrow A_{\mu a}
+\partial_\mu\lambda + A_{\mu r}\lambda_sf^{rs}_{\,\,\,\,\,\,a}
\,\,\,,
\ee
with
\be
\lambda_a = \omega_a +\theta_a\,\,\,\,,\,\,\,\,
\theta_a T^a=g^{-1}\bar\omega g\,\,\,.
\ee
A crutial property of the gauge field $A_{\mu a}$ is that its field
strength vanishes
\be
{}F_{\mu\nu a}=\partial_\mu A_{\nu a} -\partial_{\nu} A_{\mu a}
+ A_{\mu r}A_{\nu s}f^{rs}_{\,\,\,\,\,\,a}=0\,\,\,.
\ee
\par
The gauge invariant action is then given by
\be
I(g)=-{k\over 8\pi}\int_{\partial B}d^2x\sqrt{-\eta}\eta^{\mu\nu}
g^{ab}A_{\mu a}A_{\nu b}
+{ik\over 12\pi}\int_{B}d^3y\epsilon^{\mu\nu\rho}g^{ab}
(\partial_\mu A_{\nu a})A_{\rho b}
\,\,\,,
\ee
where $B$ is a three-dimensional manifold
whose boundary is the the two-dimensional surface $\partial B$.
\par
The only requirement that we have on the symmetric matrix
$g^{ab}$ is that it is an invariant of the group
$\left(f^{ab}_{\,\,\,\,\,\,c}g^{cd}+
f^{ad}_{\,\,\,\,\,\,c}g^{cb}=0\right)$. Therefore, the above
action is valid for both semi-simple and non-semi-simple groups.
Notice that the Wess-Zumino term  in this action can be
transformed into the usual form by virtue of equation (4.5).
\par
Using the fact that $F_{\mu\nu a}$ vanishes and $g^{ab}$ is
an invariant of the group, the variation of the action
is
\be
\delta I(g)=-{k\over 4\pi}\int_{\partial B}d^2x
\left(\sqrt{-\eta}\eta^{\mu\nu}-i\epsilon^{\mu\nu}\right)
g^{ab}\left[A_{\mu a}\partial_\nu \omega_b
+A_{\nu a}V^c_b\partial_{\mu}\bar\omega_c\right]\,\,\,,
\ee
where $V^a_b$ is defined through
\be
V^a_bT^b=g^{-1}T^ag
\ee
and satisfies the following equation (which was also used
to derive $\delta I$)
\be
\partial_\mu V^a_b=V^a_rA_{\mu s}f^{rs}_{\,\,\,\,\,\,b}\,\,\,.
\ee
\par
In complex coordinates $(z,\bar z)$ such that $\eta^{z\bar z}=1$
and $\epsilon^{z\bar z}=i$, we see that this variation of the action
vanishes if $\omega_a=\omega_a(z)$ and $\bar\omega_a=
\bar\omega_a(\bar z)$. The Noether current associated to some Lie
algebra element,$T^a$, are given by
\be
J^a_z=-{k \over 4\pi}g^{ab}A_{zb}\,\,\,\,,\,\,\,\,
J^a_{\bar z}=-{k \over 4\pi}g^{bc}A_{\bar z b}V^a_c\,\,\,.
\ee
The currents $J^a_z$ and $J^a_{\bar z}$ are, by virtue
of the equations of motion, holomorphic and antiholomorphic
currents, respectively. These currents do satisfy two copies
of the current algebra (2.5) as shown in [18].
\par
To summarise, we have constructed bosonic and supersymmetric
current algebras based on non-semi-simple groups.
The central charge corresponding to the Sugawara construction
is not, in general, an integer number. In fact it depends on
a free parameter. The key observation in our construction is
that the bilinear form used to define the current
algebra is taken to be different from the bilinear form
which raises and lowers group indices. We have also
constructed a WZNW action using the bilinear form entering
the operator product expansions of the current algebra.
\par
It is worth mentioning that in the case of the centrally
extended two dimensional Euclidean group studied in [1],
the difference between the two bilinear form (the
invertible one and the one used in the operator
product expansions) amounts to a difference
in the parameter $b$ defined in [1].
It is terefore of interest to explore our construction
for other non-semi-groups.

\vspace{0.5cm}
\paragraph{Acknowledgements:}
I would like to thank Lazlo {}Feh\'er, Werner Nahm
and Niels Obers
for  many useful discussions. This research
is  supported  by the Alexander von
Humboldt-Stiftung.

\end{document}